# Controlled Dephasing of Electrons by Non-Gaussian Shot Noise

*I. Neder,[1] F. Marquardt,[2] M. Heiblum,[1] D. Mahalu,[1] and V. Umansky[1]

[1]*Braun Center for Submicron Research, Department of Condensed Matter Physics,*

*Weizmann Institute of Science, Rehovot 76100, Israel*

[2]*Physics Department, Arnold Sommerfeld Center for Theoretical Physics, and Center for*

*NanoScience, Ludwig-Maximilians-Universität München, 80333 München, Germany*

## Introductory Paragraph

In a 'controlled dephasing' experiment [1-3], an interferometer loses its coherence due to entanglement with a controlled quantum system ('which path' detector). In experiments that were conducted thus far in mesoscopic systems only partial dephasing was achieved. This was due to weak interactions between many detector electrons and the interfering electron, resulting in a Gaussian phase randomizing process [4-10]. Here, we report the opposite extreme: a complete destruction of the interference via strong phase randomization only by a few electrons in the detector. The realization was based on interfering edge channels (in the integer quantum Hall effect regime, filling factor 2) in a Mach-Zehnder electronic interferometer, with an inner edge channel serving as a detector. Unexpectedly, the visibility quenched in a periodic lobe-type form as the detector current increased; namely, it periodically decreased as the detector current, and thus the detector's efficiency, increased. Moreover, the visibility had a V-shape dependence on the partitioning of the detector current, and not the expected dependence on the second moment of the shot noise, $T(1-T)$, with $T$ the partitioning. We ascribe these unexpected features to the strong detector-interferometer coupling, allowing only 1-3 electrons in the **detector to fully dephase the interfering electron. Consequently, in this work we explored the *non-Gaussian* nature of noise [11], namely, the direct effect of the shot noise full counting statistics [12-15].**

Our system is based on the previously developed electronic *two path* Mach-Zehnder interferometer (MZI) [16-18]. This time we employed two edge channels in the integer quantum Hall effect regime, at filling factor *ff*=2 (see Fig. 1). The inner edge channel was partitioned and served as a *which path* detector. Other than that, the device and the measurement technique were similar to those described in Refs. 16-18 (see also caption of Fig. 1). The MZI was fabricated within a high mobility two dimensional electron gas. The two paths were formed by splitting the outer edge channel with a quantum point contact constriction **QPC1**. After enclosing a magnetic flux the two paths joined in **QPC2** and interfered. Metallic Ohmic contacts served as sources **S1**, **S2**, and **S3** and drains **D1** and **D2**. Changing the enclosed flux by $\Delta\Phi$ (via the modulation gate, **MG**) changed the Aharonov-Bohm (AB) phase $\varphi = 2\pi\Delta\Phi/\Phi_0$ ($\Phi_0$=h/e the flux quantum) [19], leading to phase dependent transmission coefficients; say, from **S2** to **D2**:

$$T_{S2-D2} \equiv T_{MZI} = \left|t_{QPC1}t_{QPC2} + e^{i\varphi}r_{QPC1}r_{QPC2}\right|^2 = T_0 + T_\varphi \cos\varphi, \qquad (1)$$

with *t* and *r* the corresponding transmission and reflection amplitudes. The measured visibility, defined as $\nu=T_\varphi/T_0$, ranged from 30% to 60% [16-18]. We attribute the non-ideal visibility to phase fluctuations due to external noise [9].

The inner edge channel served as a path detector (see caption of Fig. 1). When **QPC0** was tuned to partition the detector channel (which was biased, $V_{det}=V_{S3}$), electrons in the upper path of the interferometer became entangled with those in the detector, resulting in a lower visibility. This dephasing process can be looked at as 'path detection' [18], or, alternatively, as phase scrambling due to potential fluctuations in the partitioned detector channel [20]. The interaction between the inner and the outer cannels was characterized before the actual dephasing experiment by first fully transmitting and then fully reflecting the biased inner edge channel emanating from **S3** (with **QPC0**). Full transmission ($T_{QPC0}$=1) did not lead to an observable effect on the AB oscillations of the MZI as function of $V_{S3}$ (Fig. 2a). However, full reflection ($R_{QPC0}$=1) had a strong effect on the phase of the interference pattern, which varied linearly with $V_{S3}$ (reaching ~$2\pi$ for $V_{det}$~19μV), but with nearly no effect on the visibility (Fig. 2b). For $V_{det}$~19μV and an interferometer dwell time $\tau = \frac{L}{v_g}$ (with $L$≈10μm, $v_g = (3-10) \cdot 10^6$ cm/sec), we estimated a mere $n$=1-3 electrons suffice to quench the interference. This strong coupling between the edges sets the present experiment apart from previous ones.

When **QPC0** was tuned to *partition* the inner channel (0<$T_{QPC0}$<1), the visibility diminished as $V_{det}$ increased. We show in Fig. 3 the dependence of the visibility on $T_{QPC0}$ (partitioning) for three different detector voltages. As the bias $V_{S3}$ increased, the visibility turned from a smooth parabolic curve to a sharp, V-shape like dependence, with a minimum at $T_{QPC0}$~0.5. The dispersion among the experimental points at higher bias resulted from resonances in $T_{QPC0}$ (see inset). We argue below that the V-shape dependence is a signature of the non-Gaussian nature of the detector noise.

We first study a simple model where exactly one electron in the detector scrambles the phase of an interfering electron. Detector electrons were injected with a probability $R_{QPC0}$=1-$T_{QPC0}$ into the channel that interacted with the interferometer. Depending on the presence or absence of a detector electron, the extra phase $\delta\varphi$ acquired by an interfering electron fluctuated between two values: $\delta\varphi = \gamma V_{det}$ ($\gamma = \frac{2\pi}{19} rad/\mu V$, from Fig. 2b) and $\delta\varphi = 0$, respectively. Averaging the $\cos\varphi$ term in Eq. (1) over the two possibilities leads to a visibility [2,13]:

$$v = \left|\left\langle e^{i\delta\varphi}\right\rangle\right| = \left|T_{QPC0} + R_{QPC0}e^{i\gamma V_{det}}\right|. \qquad (2)$$

Equation (2) does not have fitting parameters; so it can be compared directly with the experimental results. Moreover, for small $\gamma V_{det}$ the RHS of Eq. (2) can be expanded to second order: $v \approx 1 - \frac{1}{2}(\gamma V_{det})^2 T_{QPC0}(1-T_{QPC0}) \approx e^{-\frac{1}{2}(\gamma V_{det})^2 T_{QPC0}(1-T_{QPC0})}$, agreeing with the Gaussian approximation alluded to above.

The two broken lines in Fig. 3a are the predictions of Eq. (2) at detector bias $V_{det}=4\mu V$ and $9\mu V$. For the small bias the induced phase is small ($\gamma V_{det} < \pi/2$) and both Eq. (2) and the Gaussian approximation agree well with the experimental data seen in Fig. 3. However, for the larger bias the shape predicted by Eq. (2) deviates markedly from the smooth Gaussian approximation, and is a V-shape dependence: $v = \left|1 - 2T_{QPC0}\right|$ for $\gamma V_{det} = \pi$ ($V_{det} \approx 9\mu V$). In Fig. 3 this shape is indeed observed, but at a higher detector bias than that predicted by Eq. (2) ($V_{det}=14\mu V$).

Another prediction of Eq. (2) is an oscillatory dependence of the visibility on bias (the coherence should be completely recovered at $\gamma V_{det} = 2\pi n$). We plotted in Fig. 4a the dependence of the measured visibility and the average phase shift on detector bias (at $T_{QPC0}\sim 0.5$). While in Fig. 3 of Ref. 18 the visibility was found to decay monotonously with $V_{det}$, here we found, in a region of **QPC0** gate voltages which was relatively smooth and free of resonances, a non-monotonous decay. The visibility dropped to zero at $V_{det}$=14µV (instead of at 9.5µV according to Eq. 2), increased afterwards to reach another, yet smaller, maximum at $V_{det}$=22µV, and finally vanished at a higher bias. Moreover, the phase of the AB oscillations increased monotonously with $V_{det}$ (see Fig. 4b): $\langle\delta\varphi\rangle = R_{QPC0}\gamma V_{det}$, but underwent a π phase slip when the visibility reached zero; as expected qualitatively from Eq. (2).

The presently observed lobe pattern of the quenched visibility strikingly resembles the lobe-type evolution of the visibility in a self biased (by $V_{S2}$), single channel MZI [17]. This similarity suggests that intra-channel interactions (that couple an individual interfering electron to the shot noise produced by the other electrons in the same edge channel) play the role of inter-channel interactions here. A very recent theoretical preprint [21] also finds visibility oscillations in a closely related model.

To overcome the quantitative shortcomings of Eq. (2) we sketch now a more microscopic approach that predicts the main observed features. A full description of this approach will be provided in a subsequent publication [22]. We assume that every interfering electron accumulates a random phase $\delta\varphi$ as it traverses the upper arm of the MZI, due to the coupling with the fluctuating electron density in the detector channel. Treating first the detector density $\rho_{det}$ classically, the phase should be:

$$\delta\varphi = \int_0^\tau \int u(v_g t - x)\rho_{det}(x,t)\,dx\,dt = \int w(x)\rho_{det}(x)\,dx, \qquad (3)$$

with the inter-channel interaction potential $u(x)$, the electron velocity in the MZI $v_g$, and $\tau$ the traversal time in the upper path. The electron density, propagating with velocity $v_{det}$, obeys: $\rho_{det}(x,t) = \rho_{det}(x-v_{det}t,0) \equiv \rho_{det}(x-v_{det}t)$, which yields $w(x) = \int_0^\tau u[(v_g - v_{det})t' - x]dt'$. It can be shown that Eq. (3) can be used even in the quantum case ($\delta\varphi \mapsto \delta\hat{\varphi}$ and $\rho_{det} \mapsto \hat{\rho}_{det}$) to calculate the visibility $v = |\langle e^{i\delta\hat{\varphi}} \rangle|$ as long as the interfering electron is treated in a single-particle picture. This approach neglects Pauli blocking [7], which has to be taken into account in a phenomenological way while evaluating the visibility. We find (see the Methods section) that the visibility is a product of factors, each in the form of the single particle expression of Eq. (2):

$$v = |\langle e^{i\delta\hat{\varphi}} \rangle| = \prod_j |T_{QPC0} + R_{QPC0}e^{i\delta\varphi_j}| \quad . \tag{4}$$

The phases $\delta\varphi_j$ are the eigenvalues of the matrix $w_{k'k}$ (the Fourier transform of $w(x)$), that has been restricted to transitions between plane wave states $k',k$ within the voltage window. They depend on the detector voltage $V_{det}$ and obey a 'sum rule': $\sum_j \delta\varphi_j = \gamma W_{det}$. In the limit $V_{det} \to 0$ one can show that only one nonzero eigenvalue remains and the result reduces to Eq. (2) [22].

Choosing the Fourier transform of $w(x)$ as a Lorentzian, with its FWHM as the single fitting parameter ($\Delta$=12.4μeV; the height being deduced from the observed value of $\gamma$), we plotted in Fig. 4 the calculated visibility from Eq. (4). The plot reproduced the phase slip and zero visibility at $V_{det}$=14μV, the second lobe, and the eventual decay at higher detector voltages. Since more than one detector electron participated in the dephasing process, the largest eigenvalue $\delta\varphi_1$ becomes smaller than $\gamma W_{det}$ (because of the 'sum rule'), such that the zero visibility (when $\delta\varphi_1 = \pi$) is reached at a higher $V_{det}$ than predicted by Eq. (2). Though the quality of the fit is rather good, it may be even further improved if $w(x)$, determined by the microscopic physics of the edge channels, were known more precisely. We may conclude that for $V_{det}$<6μV a single detecting electron dephases the MZI, while at $V_{det}$~30μV the number is at most three.

In summary, we presented a unique behavior of an electron interferometer coupled to a *which path* detector. Very strong interactions between electrons in both systems led to dephasing by the characteristic binomial, non-Gaussian, shot noise in the detector. The dephased visibility had a linear, V-like shape dependence on the partitioning of the detector's current, and non-monotonic, periodic, lobe pattern decay as a function of the detector current itself. This entanglement between nearly single pairs of electrons may be exploited (in future experiments) to test Bell's inequalities in a system where the detector channel is replaced by another two-path interferometer [23-25].

**Acknowledgement**

We are indebted to Y. Levinson for helpful discussions. The work was partly supported by the Israeli Science Foundation (ISF), the Minerva foundation, the German Israeli Foundation (GIF), the SFB 631 of the DFG, and the German Israeli Project cooperation (DIP).

**Competing financial interests**

The authors declare that they have no competing financial interests.

**Methods**

**Sample and measurements**

The edges of the sample are defined by plasma etching of a GaAs-AlGaAs heterostructure, embedding a high mobility 2D electron gas, 80nm below the surface. Two edge channels are formed by applying a perpendicular magnetic field of ~3T, leading to a filling factor 2 in the bulk (electron temperature ~15mK). Transmission of the outer channel to **D2** is measured by applying ~1μV at ~1MHz at **S2**. The signal at **D2** (see Fig. 1), filtered by a cold LC resonant circuit tuned to 1MHz with bandwidth 30kHz, is amplified by a low noise preamplifier at 4.2K. Note that the inner, small, Ohmic contact (3x3μm$^2$) serves as both **D1** and **S1**.

**Evaluation of visibility**

Being a many-particle quantum device, the detector's density cannot be expressed either as a classical function or in a single particle language. Hence, Eq. (3) should be rewritten in terms of detector electron operators $\hat{d}_k$ and the matrix elements of $w(x)$ with respect to a plane wave basis:

$$\delta\hat{\varphi} = \sum_{k,k'} w_{k'k} \hat{d}^+_{k'} \hat{d}_k . \qquad (1')$$

The occupation of each $k$-state fluctuates independently with $n_{k'k} \equiv \langle \hat{d}^+_{k'} \hat{d}_k \rangle = n_k \delta_{k'k}$, with $n_k = 1/R_{QPC0}/0$ (at zero temperature) for $k$ below/within/above the detector voltage window $E_F<E(k)<E_F+eV_{det}$. It now becomes possible to express the expectation value of the many-body operator $e^{i\delta\hat{\varphi}}$ in terms of a determinant involving the matrices $w_{k'k}$ and $n_{k'k}$ [13]:

$$\langle e^{i\delta\hat{\varphi}} \rangle = det\left[1 + (e^{iw} - 1)n\right] . \qquad (2')$$

Equation (2') can be evaluated numerically. It converges in the limit of a large normalization volume and large upper/lower cutoffs in $k$. However, this expression leads to a suppressed visibility even at zero temperature and $V_{det} = 0$, which is an artifact of neglecting the Pauli blocking (which prevents the interfering electrons from scattering into occupied states below $E_F$). This can be cured approximately either by rescaling the visibility by a factor independent of $V_{det}$ (setting it to 1 at $V_{det} = 0$), or by restricting the matrix elements $w_{k'k}$ to transitions only within the detector voltage window. The latter approach allows a further simplification of Eq. (2') and yields Eq. (4) of the main text.

**Figure Captions**

**Fig. 1. The configuration of the Mach-Zehnder interferometer and the detector. a**. Schematic description of the MZI. **b.** SEM micrograph of the fabricated structure. Note the air bridges for the inner contact (**D1**) and for the two QPCs. The two channels, injected from each source, **S2** and **S3,** propagated toward **QPC0**, which was tuned to fully transmit the outer channel but partly transmit the inner one. Consequently, two channels impinged on **QPC1** from the right: a full outer (interferometer) channel (from **S2**) and a partitioned inner 'detector' channel (from **S2** and **S3**). **QPC1** and **QPC2** fully reflected the inner channel and partly transmitted the outer channel (generally $T=R=0.5$). The presence of an electron in the upper path of the interferometer affected the phase of electrons in the detector channel, and vice versa, by Coulomb repulsion.

**Fig. 2. The effect of the inner ('detector') edge channel (injected by S3) on the phase and amplitude (insets) of the AB oscillations of the outer (interferometer) channel. a**. The inner channel is fully transmitted by **QPC0** ($T_{QPC0}=1$). The phase and visibility are not affected by biasing **S3**. **b.** The inner channel is fully reflected by **QPC0** ($R_{QPC0}=1- T_{QPC0}=1$), and flows parallel and in close proximity to the outer channel upper path. The phase is highly sensitive to the bias $V_{det}$ on **S3**, $\frac{d\varphi}{dV_{det}} \cong \frac{2\pi}{19\mu V}$, while the visibility remains almost constant.

**Fig. 3. The effect of partitioning the detector channel (by QPC0) on the visibility of the interfering signal, at three different detector bias values.** As $V_{det}$ increases, the dependence of the visibility on $T_{QPC0}$ turns from a smooth one to a sharp V-shape (at $V_{det}=14\mu V$). The dashed line is the prediction of a single-detector-electron model (Eq. (2)). While the model agrees with the experimental results at low bias (4µV, black line), it fails at larger bias. It indeed predicts a V-shape dependence, but at a lower bias $V_{det}=9\mu V$ (gray line). **Inset**: The conductance of **QPC0** as function of gate voltage shows

sharp resonances. This explains the lack of visibility measurements in the range $0.1<T_{QPC0}<0.4$, and its dispersion at large detector bias (due to the dependence of the resonances on bias).

**Fig. 4. The evolution of the visibility and phase as a function of detector voltage $V_{det}$, for a partitioned detector channel.** The non-monotonic behavior of the visibility is a clear sign of dephasing by non-Gaussian noise (solid black line). Dashed line: Prediction of the improved theoretical model, which takes into account fully the effects of binomial shot noise, see text, Eq. (4). The discrepancy at negative $V_{det}$ results from slight non-linearity of **QPC0**, leading to a non-accurate $R_{QPC0}$ ($T_{QPC0}{\sim}0.5$, $V_{QPC0}{=}-0.0272$V - see inset of Fig. 3).

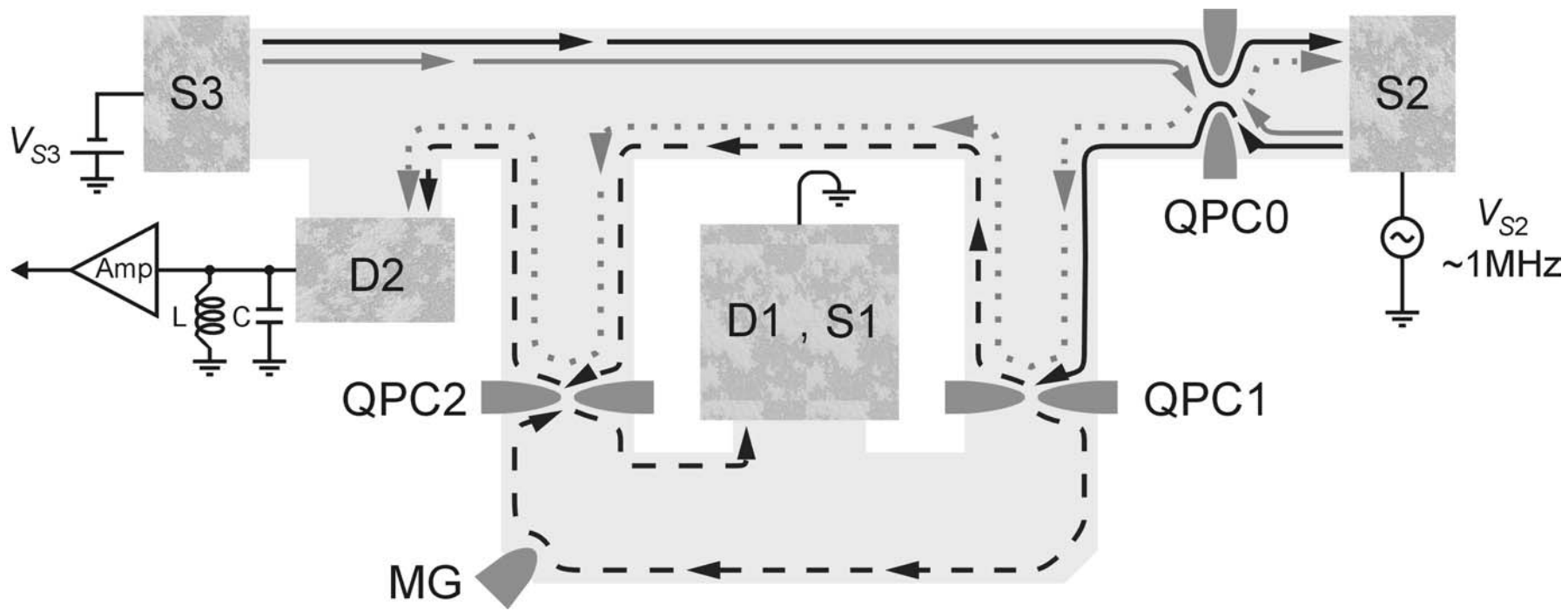

Fig. 1a

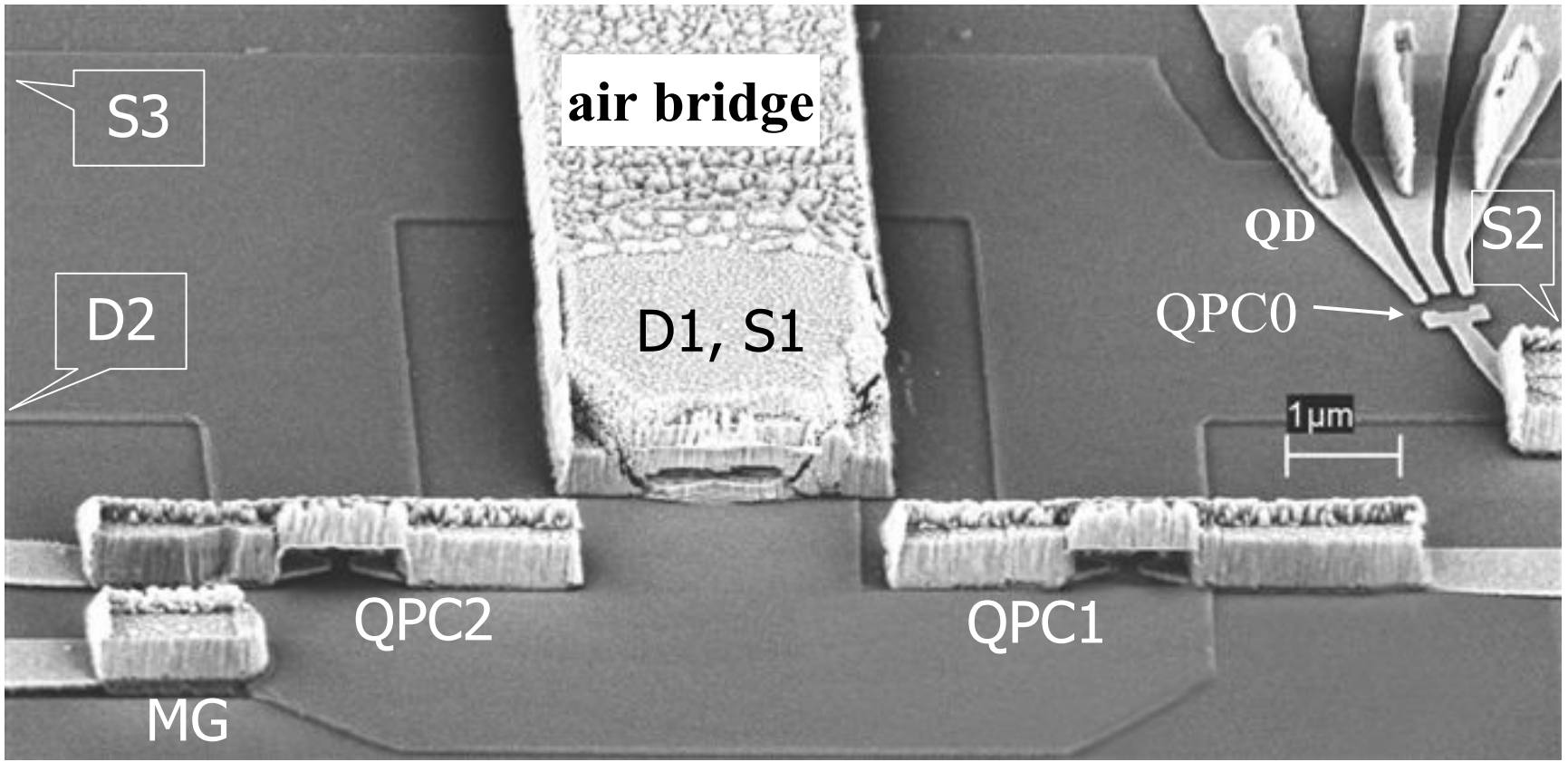

Fig. 1b

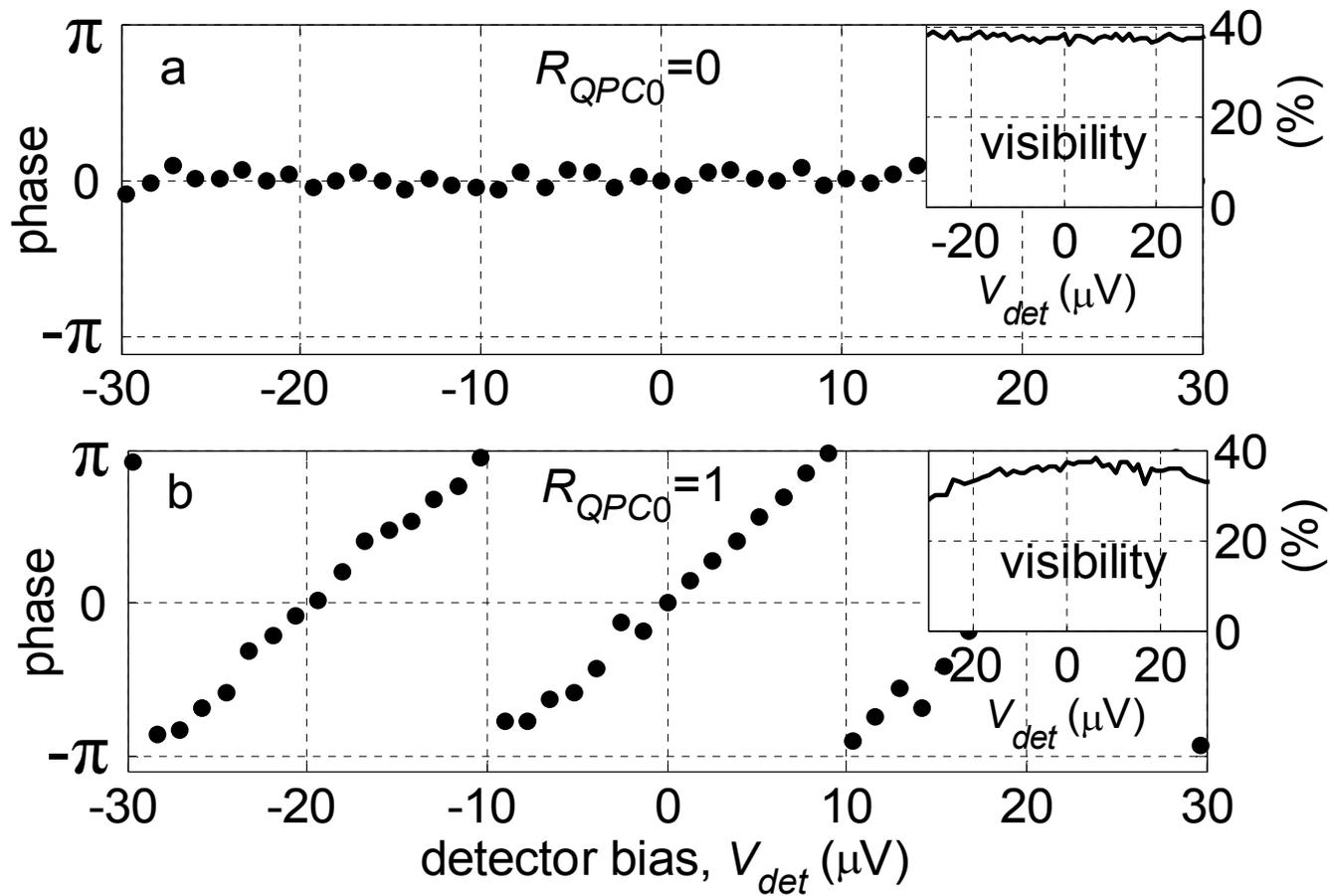

Fig. 2

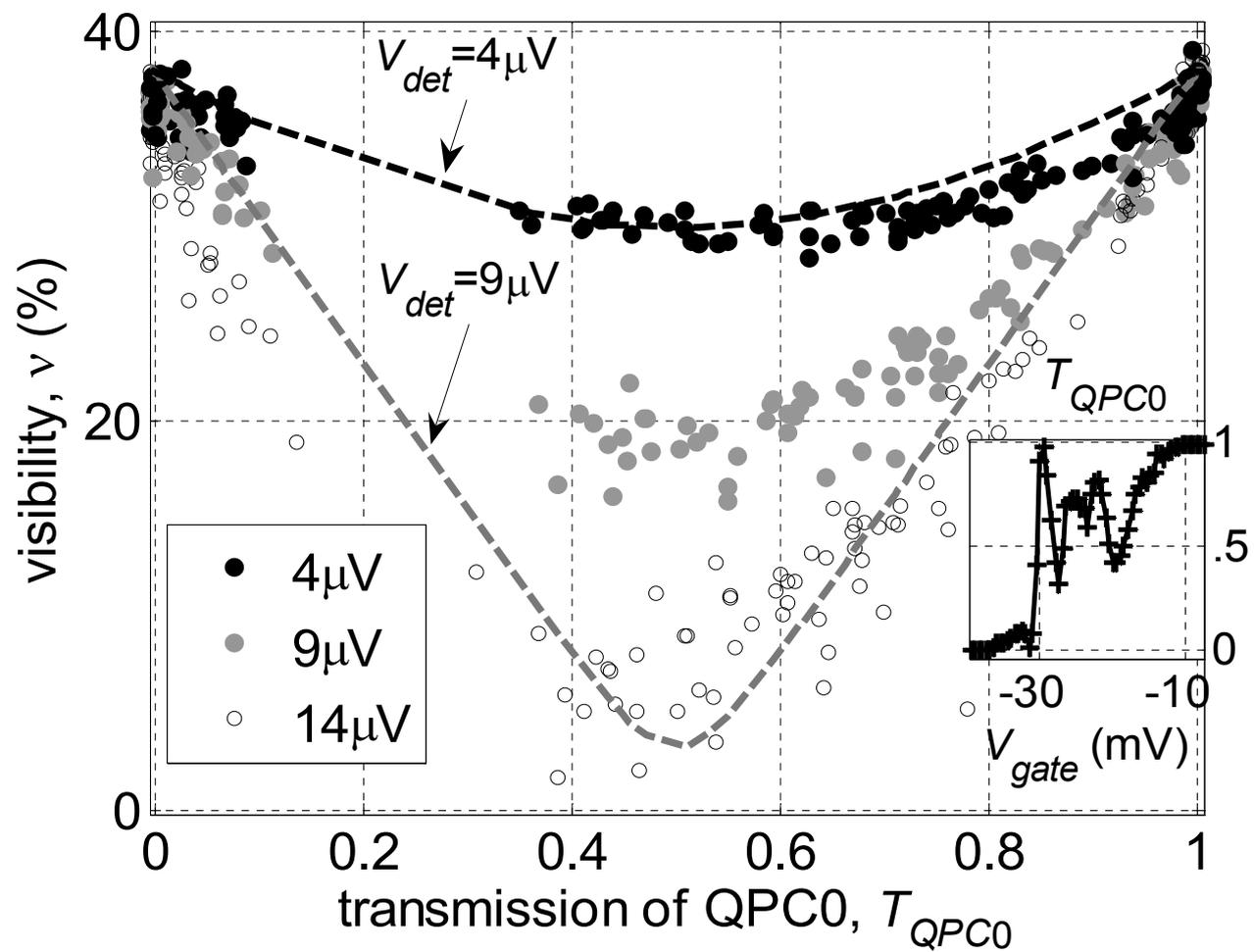

Fig. 3

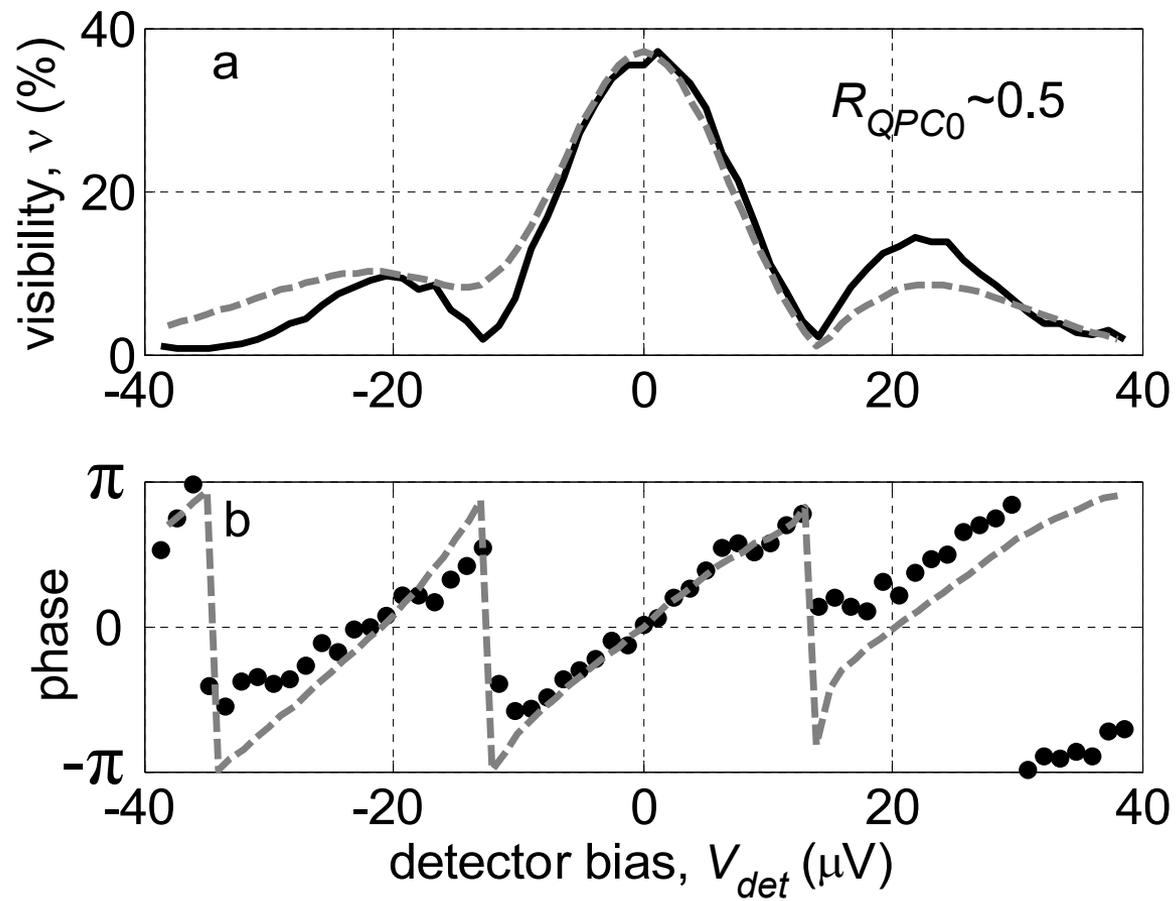

Fig. 4